\documentclass[aps,prb,reprint,amsmath,amssymb, superscriptaddress]{revtex4-2}
\usepackage{amsmath}
\usepackage{bm}
\usepackage{graphicx}
\usepackage{physics}
\usepackage{color}
\usepackage{braket}
\usepackage{soul}

\usepackage[colorlinks=true, citecolor=blue, linkcolor=blue, urlcolor=blue]{hyperref}

\begin{document}

\title{First-principles cumulant approach to the vibronic structure of spin defects}

\author{Jinsoo Park}
\email[Corresponding author: ]{js.park@postech.ac.kr}
\affiliation{Pritzker School of Molecular Engineering, University of Chicago, Chicago, IL 60637, USA}
\affiliation{Department of Physics, Pohang University of Science and Technology, Pohang 37673, Korea}
\affiliation{Institute for Theoretical Science, POSTECH, Pohang 37673, Korea }

\author{Yu Jin}
\affiliation{Pritzker School of Molecular Engineering, University of Chicago, Chicago, IL 60637, USA}
\affiliation{Initiative for Computational Catalysis, Flatiron Institute, New York, NY 10010, USA}

\author{Arpan Kundu}
\affiliation{Pritzker School of Molecular Engineering, University of Chicago, Chicago, IL 60637, USA}

\author{Jorge O. Sofo}\email[Corresponding author: ]{sofo@psu.edu}
\affiliation{Department of Physics and Materials Research Institute, The Pennsylvania State University,
University Park, Pennsylvania 16802, USA}

\author{Giulia Galli}\email[Corresponding author: ]{gagalli@uchicago.edu}
\affiliation{Pritzker School of Molecular Engineering, University of Chicago, Chicago, IL 60637, USA}
\affiliation{Materials Science Division and Center for Molecular Engineering, Argonne National Laboratory, Lemont, IL 60439, USA}
\affiliation{Department of Chemistry, University of Chicago, Chicago, IL 60637, USA}
\begin{abstract}

Color centers in wide-band-gap semiconductors are leading platforms for solid-state quantum technologies, yet a quantitative description of their vibronic structure has remained elusive due to the complexity of multi-phonon processes in localized defect states. Here we present a first-principles Green's function framework based on the retarded cumulant ansatz (RCA) to describe electron-phonon interactions in spin defects; our approach goes beyond the adiabatic and lowest-order perturbation theory approximations underlying widely used approaches. Applied to the negatively charged nitrogen-vacancy (NV$^{-}$) center in diamond, our method reveals that multi-phonon satellites persist over a 400 meV energy window even at zero temperature, driven by quantum zero-point fluctuations. We demonstrate that accurate spectral functions require mode-, momentum-, spin-, and orbital-resolved electron-phonon matrix elements sampled across the full Brillouin zone, to account for hybridized and propagating phonon channels. We find that the vibronic structure of the NV$^-$ center exhibits strong spin and orbital anisotropy, with different orbitals  coupling to qualitatively distinct parts of the phonon spectrum, and  spin-selective coupling affecting both sideband positions and intensities.

\end{abstract}
\maketitle
Color centers in wide band gap semiconductors are leading spin qubit platforms for modern quantum technologies~\cite{wolfowiczQuantum2021,liComputation2026}.
Their electronic levels are energetically isolated from the host bands and enable robust spin-qubit initialization, control, and readout by optical or magnetic probes~\cite{dohertyNitrogenvacancy2013}.
Understanding the dynamical behavior of spin qubits requires a detailed description of the coupling between their electronic, magnetic, and vibrational degrees of freedom.
For example, the electron-phonon ($e$-ph) interaction governs processes such as intersystem crossing~\cite{thieringInitio2017,jinFirstPrinciples2025}, photoluminescence~\cite{alkauskasFirstprinciples2014,jinPhotoluminescence2021}, Jahn--Teller effects~\cite{jinVibrationally2022,kunduQuantum2024}, and spin relaxation~\cite{parkSpinphonon2020,parkManybody2022,parkPredicting2022,cambriaTemperatureDependent2023}.

Several frameworks have been proposed to describe the $e$-ph interaction in spin  defects.
The Huang-Rhys (HR) theory uses the displaced harmonic oscillator approximation to evaluate lineshapes via a generating function~\cite{alkauskasFirstprinciples2014,jinPhotoluminescence2021}, and has successfully been used to investigate the photoluminescence lineshapes and photoionization cross sections~\cite{razinkovasVibrational2021} of several systems.
In addition, a stochastic approach~\cite{monserratVibrational2016, kunduQuantum2023,kunduQuantum2024,kundu_nanodiamond} has been adopted, that samples vertical excitation energies over thermally distributed atomic geometries.
This method offers the advantage of including the anharmonicity of the ground state potential energy surface, making it particularly suitable to study the electronic structure of  defects  at finite temperature.

Despite the success of the HR theory and of stochastic approaches, these frameworks have several limitations.
Both of them  rely on the adiabatic approximation and the evaluation of static potential energy surfaces, and do not include the full mode- and momentum-resolved $e$-ph interaction of the defect states, nor the dynamical properties of the electron self-energy.
Further, the stochastic approach neglects the dynamical correlations inherent in the time-ordered sequence of atomic vibrations.
Non-adiabatic self-energy corrections to defect energy levels were investigated in Refs.~\cite{yangCombined2021,yangComputational2022} using density-matrix perturbation theory, demonstrating improved accuracy relative to conventional treatments based on the adiabatic approximation. However, the approach of Refs. ~\cite{yangCombined2021,yangComputational2022} does not account for multi-phonon satellite features associated with the repeated emission and absorption of phonons.

Here we present a method to compute electron–phonon interactions in spin defects by combining mode-resolved $e$–ph coupling with a dynamical many-body treatment of localized defect states. Electron–phonon matrix elements are evaluated using a localized-orbital representation together with Wannier interpolation~\cite{giustinoElectronphonon2007,agapitoInitio2018,zhouPerturbo2021}, enabling calculations in large supercells. Our approach is based on the retarded cumulant ansatz (RCA), previously successfully applied to describe satellite features in
X-ray photoemission of core electrons, electrons interacting with plasmons~\cite{langrethSingularities1970}, or phonons~\cite{kasCumulant2014,storyCumulant2014,neryQuasiparticles2018}.
  Green's-function methods have enabled parameter-free predictions
 of a wide range of dynamical properties in solids, including electron scattering~\cite{zhouInitio2016}, charge transport~\cite{sohierPhononlimited2014,liElectrical2015,zhouElectronPhonon2018}, ultrafast dynamics~\cite{maliyovInitio2021}, and spin relaxation and decoherence~\cite{parkSpinphonon2020,parkManybody2022,parkPredicting2022}. However, these approaches have so far been largely restricted to extended states. Using the negatively charged nitrogen-vacancy (NV$^-$) center in diamond as a prototypical spin defect, we demonstrate that a Green's-function-based framework accurately captures the effects of multi-phonon excitations on localized defect states, including satellite structures that are beyond the reach of lowest-order perturbation theory.

The retarded Green's function for a given defect state $n$ is
\begin{equation}
G_n^R(t)=-i\theta(t)\langle \{c_n(t),c_n^\dagger(0)\}\rangle,
\end{equation}
where $c_n^\dagger$ and  $c_n$ are the creation and annihilation operators, respectively.
 The retarded cumulant $C_n^R(t)$ ~\cite{kasCumulant2014,storyCumulant2014,zhouPredicting2019,neryQuasiparticles2018} is expressed in terms of the imaginary part of a retarded self-energy $\Im \Sigma_n^R$,
\begin{equation} \label{eq:cumulant}
   C_n^R(t)=\int_{-\infty}^{\infty} d\omega \frac{\beta_n(\omega)}{\omega^2}(e^{-i\omega t}+i\omega t -1)
\end{equation}
where $\beta_n(\omega)=|\Im \Sigma_n^R(\omega+\varepsilon_n)|/\pi$, and $\varepsilon_n$ is the Kohn-Sham single particle energy.
We compute $\beta_n(\omega)$ by expressing the $e$-ph interaction with the lowest-order Fan-Migdal self-energy $\Sigma_{\mathrm{e-ph}}$, and the  electron-electron interaction at the   $\mathrm{G_0W_0}$ level of theory.
Note that the cumulants of defect states are independent of the wave vector because of their spatial localization.
The retarded Green's functions are computed using the retarded cumulant,
\begin{equation}\label{eq:G}
G_n^R(t)=-i\theta(t)e^{-i\varepsilon_n t }e^{C_n^R(t) }.
\end{equation}
This expression shows that the energy shift produced by the environment on the electronic energy is given by the imaginary part of the cumulant and its broadening is given by its real part.
The spectral functions from the RCA are obtained from the imaginary part of the Fourier transform of Eq.~(\ref{eq:G}),
\begin{equation}\label{eq:spectral_cumulant}
    A_n(\omega)=-\text{Im} G_n^R(\omega)/\pi.
\end{equation}
The RCA gives an exact Green's function of an isolated electron dressed by harmonic bosons~\cite{langrethSingularities1970,nessApproximations2011} in the independent boson model~\cite{mahanManyParticle2000} (see Supplemental Material~\cite{supp_mat_prb}), indicating that
 the RCA is an appropriate framework to describe localized defect levels dressed by interactions with the environment.

Using the {\sc Quantum ESPRESSO}~\cite{giannozziQUANTUM2009} code, we computed the ground state electronic structure and phonon dispersion of a NV$^-$ center in diamond with DFT and the PBE exchange-correlation functional~\cite{perdewGeneralized1996}, and a 511-atom supercell  with a lattice constant of 3.57~$\mathring{\text{A}}$. The energy cutoff for the electronic wavefunction is 60 Ry.
We used the {\sc perturbo} code~\cite{zhouPerturbo2021} to compute and interpolate the $e$-ph matrix elements using Wannier functions generated by {\sc Wannier90}~\cite{pizziWannier902020}.
The self-energies and the spectral functions were evaluated using a converged $40\times40\times40$ phonon momentum grid.
We employed the  WEST~\cite{govoniLarge2015,yuGPU2022} code to compute the full-frequency $\mathrm{G_0W_0}$ self-energy~\cite{rohlfingElectronhole2000,scherpelzImplementation2016,govoniGW1002018,maFiniteField2019} of the defect states, with 4096 projective dielectric eigenpotentials~\cite{nguyenImproving2012,phamCalculations2013,wilsonEfficient2008,wilsonIterative2009}, and 6~Ry energy cutoff for the real frequency integration; we used a 0.04~$-$ 0.4~eV energy broadening, and a Lanczos chain~\cite{walkerEfficient2006,roccaTurbo2008} constructed with 100 elements to ensure convergence.
\begin{figure*}[t!]
\includegraphics[width=1.0\textwidth]{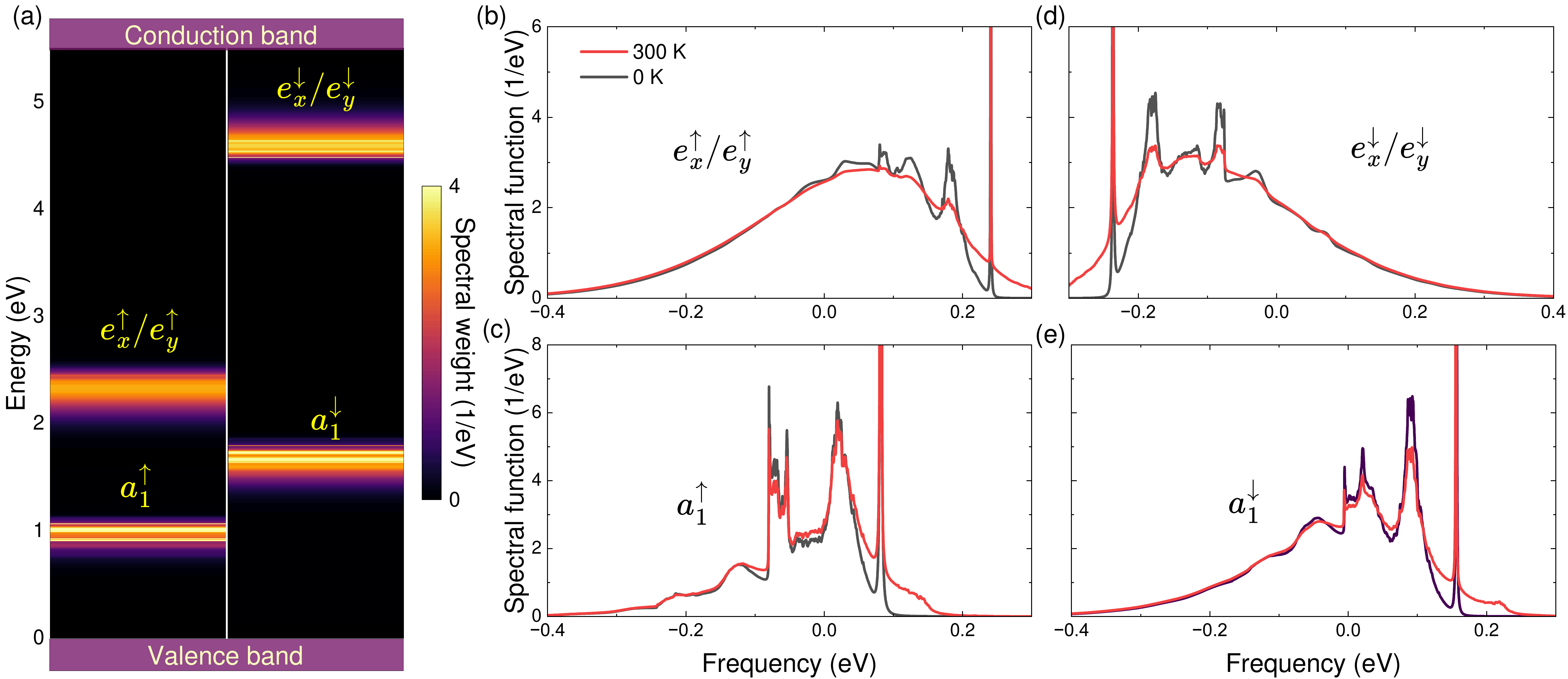}
\caption{Electron spectral functions of the defect states.
(a) Computed spectral weight of the defect states in the triplet ground state of $\rm NV^-$ center in diamond at 300~K using the retarded cumulant ansatz.
The energy is referenced to the $\mathrm{G_0W_0}$ valence band maximum.
(b)-(e) Spectral functions of the (b) $e_x^\uparrow$ and $e_y^\uparrow$ states, (c) $a_1^\uparrow$ state, (d) $e_x^\downarrow$ and $e_y^\downarrow$ states, and (e) $a_1^\downarrow$ state.
Shown are the computed results from the retarded cumulant ansatz at 300~K (solid red line) and 0~K (solid black line). The zero of frequency corresponds to the $\mathrm{G_0W_0}$ quasiparticle (QP) energy.
}\label{fig:spectra}
\end{figure*}

Figure~\ref{fig:spectra}(a) shows the ground state electron spectral weight $\sum_n A_n(\omega)$ of the defect states associated with the $\rm NV^-$, as computed  with the RCA.
The spectral weight is notably broad, with a full-width half-maximum of the sidebands of the $e_x$ and $e_y$ states of $\sim$400~meV at 0~K.
The spectral weight exhibits replicas of the quasiparticle (QP) peaks separated by energies of 60--170~meV, which are most clearly visible for the $a_1^\uparrow$ and $a_1^\downarrow$ states.
We note that this broad sideband spectrum arises when summing over all phonon modes, as done when using the RCA; it does not appear in the spectral weight computed from the lowest-order Fan-Migdal self-energy which yields instead a Lorentzian lineshape with  a narrow broadening of less than 1~meV (see Supplemental Material~\cite{supp_mat_prb}).
Overall, our results show that, as expected, even at 0~K zero-point vibronic coupling strongly affects the energy of the defect levels; in addition they show that the spectral weights extend over a broad energy window rather than yielding a single sharp quasiparticle peak.

Figures~\ref{fig:spectra}(b)-(e) display the temperature-dependent spectral function $A_n(\omega)$ of the six defect states within the band gap of diamond. All spectral functions exhibit a QP peak and multiple sidebands originating from the interaction with phonons (phonon sidebands), but interestingly they each show unique features.

The positions of the QP peaks are shifted from the  $\mathrm{G_0W_0}$ value computed at $\omega=0$. These shifts originate from the finite real part of the Fan-Migdal self-energy.
Shifts are positive for the occupied $a_1^\uparrow$, $e_x^\uparrow$, $e_y^\uparrow$, $a_1^\downarrow$ states, and negative for the unoccupied $e_x^\downarrow$ and $e_y^\downarrow$ states.
As a result, the QP energy gap between the highest occupied and the lowest unoccupied defect state decreases,
an effect analogous to that observed for the band gap renormalization of solids due to phonons~\cite{mahanManyParticle2000,lihmPhononinduced2020}.

The phonon sidebands of all six states are significantly broader compared to the respective QP peaks, and their position, relative to the QP peak, depends on the electron occupation factor.
The sidebands of the occupied states (Figures~\ref{fig:spectra}(b), (c), and (e)) primarily extend  at energies lower than that of the QP peak, due to the excitation of phonons during electron emission processes.
Instead, for the unoccupied $e_x^\downarrow$ and $e_y^\downarrow$ states (Fig.~\ref{fig:spectra}(d)), the sidebands lie mainly above the QP peak, arising from phonon shake-up effects during inverse photoemission processes.
Specifically, at 0~K, all phonon sidebands lie entirely below or above the QP peak.
At 300~K, a small fraction of the spectral weight emerges on the opposite side of the QP peak, due to thermally excited phonons, that can annihilate and assist in the electron photoemission process.

We find that, as electrons or holes continuously emit and absorb phonons, the combined energy fluctuations of the QP peak and phonon sidebands are close to their value in the absence of $e$-ph interaction. This result is consistent with the harmonic approximation adopted here. Indeed,
in the independent boson model, the mean position of the energy, $\langle\omega\rangle_n=\int \omega A_n(\omega)d\omega $, is averaged to zero and coincides with that of the undressed electron energy~\cite{mahanManyParticle2000}.
As a consequence, the time-averaged energy levels are identical to their respective energies in the absence of $e$-ph interaction.

\begin{figure*}[t!]
\includegraphics[width=1.0\textwidth]{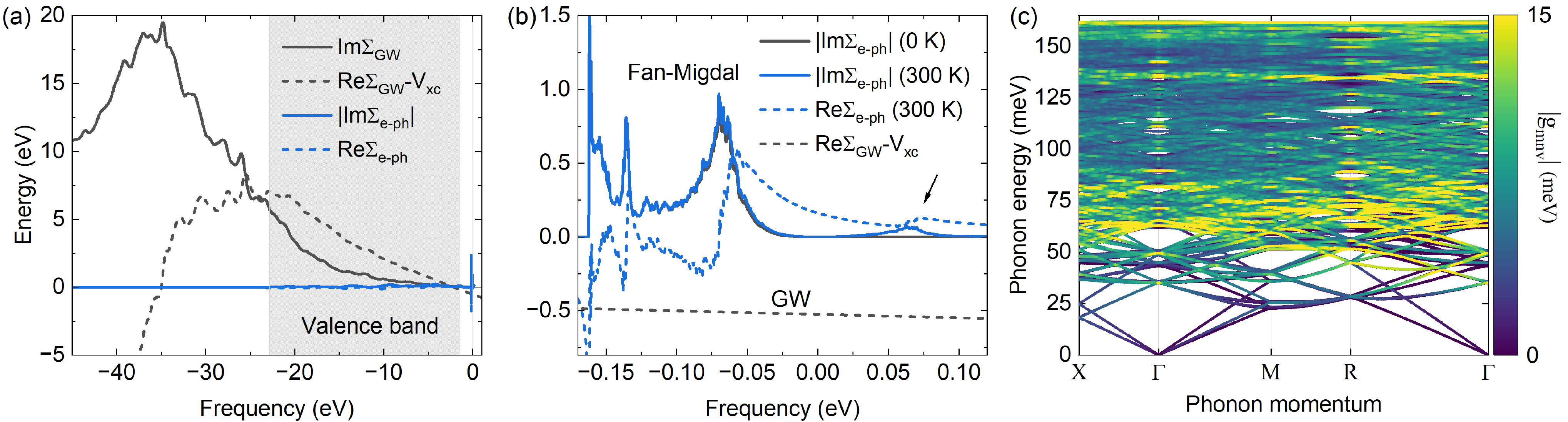}
\caption{Self-energies of the $a_1^\downarrow$ defect state.
(a) Frequency-dependent $\mathrm{G_0W_0}$ and Fan-Migdal self-energy of the $a_1^\downarrow$ defect state of $\rm NV^-$ in diamond. Shown are the real part (solid line) and the imaginary part (dotted line) of the $\mathrm{G_0W_0}$ (black) and Fan-Migdal (blue) self-energy at 300~K.
The gray area shows the valence band energy range of diamond spanned by $sp^3$ orbitals. The frequency is referenced to the Kohn-Sham eigenvalue.
(b) Zoom-in of the self-energies in (a). Imaginary part of the Fan-Migdal self-energy at 0~K (black solid line) is shown for comparison. The arrow indicates the Fan-Migdal self-energy from phonon absorption at 300~K.
(c) Phonon dispersion of the 511-atom supercell $\rm NV^-$ center in diamond, color-coded by the mode-resolved $e$-ph matrix element $|g_{q}|$ between the $a_1^\downarrow$ states.
}\label{fig:a1}
\end{figure*}
The most striking features shown in Figs.~\ref{fig:spectra}(a)-(e) are the presence of multiple peaks in the sidebands within an energy range of about 170~meV, relative to the QP frequency.
These peaks are most prominent at 0~K, where the phonon occupation factor is zero and phonon emission effects are dominant.
At higher temperatures, in addition to emission, phonon absorption occurs, and the sideband peaks are less prominent across the whole frequency range.
For instance, the sideband spectra of $e_x^\uparrow$ and $e_y^\uparrow$ states in Fig.~\ref{fig:spectra}(b) show sharp peaks at 70~meV and 170~meV below the QP peak at 0~K.
These peaks are broadened at 300~K and the spectral function evolves into a smooth Poisson-like distribution.
Note that the sharp spectral features for the $a_1^\uparrow$ and $a_1^\downarrow$ states differ in energy and intensity from those of the $e_x^\uparrow$ and $e_y^\uparrow$ states.

To understand the relative effect on the energy levels of the $e$-ph and
$\mathrm{G_0W_0}$ self-energies,  we focus on the $a_1^\downarrow$ defect state as an example.
Figure~\ref{fig:a1}(a) shows the electron self-energy from the electron-electron interaction of the $a_1^\downarrow$ state.
The strong frequency dependence of $\Sigma_{\mathrm{GW}}(\omega)$ occurs far outside the phonon energy window relevant to the in-gap defect states.
On an expanded scale in Fig.~\ref{fig:a1}(b),
$\Im\Sigma_\mathrm{GW}(\omega)$ is negligibly small, and
$\Re(\Sigma_\mathrm{GW}(\omega)-V_{xc})$ is nearly constant at $-0.5$~eV, as expected for in-gap defect states with suppressed low-energy electronic decay channels.
In contrast, the Fan-Migdal self-energy $\Sigma_{\mathrm{e-ph}}(\omega)$ has a pronounced structure with a maximum of $|\Im\Sigma_{\mathrm{e-ph}}(\omega)|\approx2$~eV near $\omega=-170$~meV associated with optical phonon emission.
Consequently, the lineshapes are governed primarily by $\Sigma_{\mathrm{e-ph}}(\omega)$.
We conclude that the spectral functions of the color center defect states can be accurately described by a static $\mathrm{G_0W_0}$ correction combined with the full frequency-dependent $e$-ph self-energy.

Figure~\ref{fig:a1}(b) displays the imaginary part of the Fan-Migdal self-energy of $a_1^\downarrow$ defect state.
At 0~K, it is entirely below zero ($\omega < 0$), reflecting the dominant phonon emission processes.
At 300~K, thermally populated phonons enable absorption processes, creating a distinctive feature near $\omega=70$~meV and enhancing the intensity below zero.
This absorption channel generates the anti-Stokes spectral weight in Fig.~\ref{fig:spectra}(e) that appears on the opposite side of the QP peak.
The overall profile remains highly asymmetric even at room temperature, indicating that at 300~K thermal excitations are insufficient to change the physical picture obtained at 0~K, arising from quantum fluctuations (zero-point motion).

The momentum dependence of the phonons reveals whether the dominant $e$-ph interactions arise from localized vibrations or from delocalized, propagating lattice modes.
Previous works focused on $\Gamma$-point phonons~\cite{alkauskasFirstprinciples2014,jinPhotoluminescence2021,jinVibrationally2022}, and captured only a subset of vibrational modes restricted by the $C_{3v}$ crystal symmetry of the NV$^-$ center.
Here we analyze the phonon momentum dependence of the spectral intensity $\Im\Sigma_{\mathrm{e-ph}}(\omega)$ using the full phonon dispersion of the $\rm NV^-$ in Fig.~\ref{fig:a1}(c).

Figure~\ref{fig:a1}(c) shows the phonon dispersion of the 511-atom diamond supercell containing an $\rm NV^-$ center, color-coded with the $e$-ph matrix element magnitude $|g_{q}|$ for  the $a_1^\downarrow$ defect states.
Peaks in $\Im\Sigma_{\mathrm{e-ph}}(\omega)$ identify the phonon energies that dominate the scattering processes, while the momentum dependence of $|g_{q}|$ indicates whether the phonons are localized, hybridized, or propagating.

The main peak of $\Im\Sigma_{\mathrm{e-ph}}(\omega)$ occurs around $\omega=-$170~meV and stems from a high-energy optical phonon with weak momentum dependence (a near-flat phonon energy and interaction strength indicates a localized vibrational mode~\cite{jinPhotoluminescence2021}).
The feature around $\omega=-130$~meV involves multiple phonon modes; the strength is primarily focused at the $R$ point in the Brillouin zone (BZ), indicating hybridization between localized and bulk phonon modes.
The broader maximum near $\omega=-60$~meV arises from dispersive phonon energies, with mode-dependent interaction strengths, indicating an interaction with several delocalized phonon modes.
For $|\omega|<25$~meV, $\Im\Sigma_{\mathrm{e-ph}}(\omega)$ exhibits a cubic dependence on energy. Such a trend stems from the three propagating acoustic phonon modes in Fig.~\ref{fig:a1}(c).
Overall our results show that the sideband spectrum encodes a number of varied phonon channels: localized, hybridized, and propagating modes; hence a detailed description of the $e$-ph interaction of defect states requires sampling the full BZ, rather than relying on the $\Gamma$-point approximation.

\begin{figure}[t!]
\includegraphics[width=1.0\columnwidth]{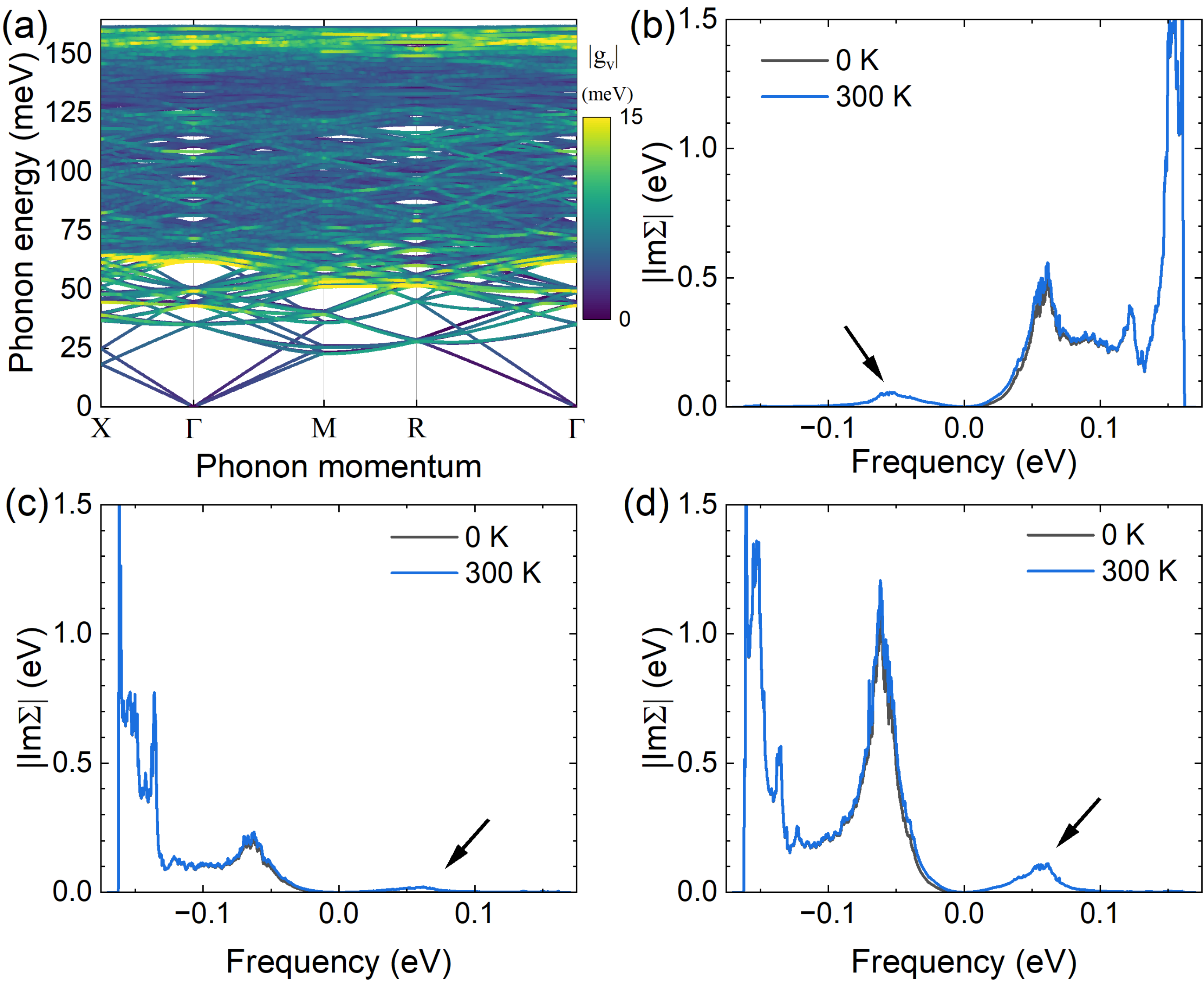}
\caption{Electron-phonon interactions and self-energies of selected NV$^-$ defect states.
(a) Phonon dispersion of the 511-atom supercell $\rm NV^-$ center. The dispersion is color-coded by the $e$-ph matrix element $|g_q|$ summed over the bands spanned by the $e_x^\downarrow$ and $e_y^\downarrow$ states.
(b)-(d) Imaginary part of the Fan-Migdal self-energy of the (b) unoccupied $e_x^\downarrow$ and $e_y^\downarrow$ states, (c) occupied $e_x^\uparrow$ and $e_y^\uparrow$ states, and (d) occupied $a_1^\uparrow$ state. Shown are the computed results
at 0~K (black) and 300~K (blue).  Arrows indicate the self-energy from phonon absorption.
}\label{fig:exey}
\end{figure}

The $e$-ph coupling strength exhibits orbital and spin anisotropy, as
shown by the state-resolved $e$–ph matrix elements and Fan–Migdal self-energy for the  defect levels.
Figure~\ref{fig:exey}(a) shows the phonon dispersion of $\rm NV^-$ color-coded with an averaged $e$-ph interaction strength $|g_q|=\sqrt{\sum_{nm}|g_{nmq}|^2/2}$ for the  $e_x^\downarrow/e_y^\downarrow$ manifold, where the band indices $n$ and $m$ sum over the $e_x^\downarrow$ and $e_y^\downarrow$ states.
The interaction strength is concentrated near $\omega=60$~meV and $\omega=160$~meV. Correspondingly, $\Im\Sigma_{\mathrm{e-ph}}(\omega)$ exhibits peaks at these two frequencies in Fig.~\ref{fig:exey}(b).
Since these states are unoccupied, the self-energy at 0~K is non-zero only for positive frequencies ($\omega>0$), corresponding to phonon emission processes. At 300~K, a small thermal absorption tail appears at negative frequencies.

Spin polarization is another important property to take into account when computing the $e$-ph coupling.
Figure~\ref{fig:exey}(c) shows $\Im\Sigma_{\mathrm{e-ph}}(\omega)$ for the occupied $e_x^\uparrow$ and $e_y^\uparrow$ states.
Comparing the $e_x$ and $e_y$ self-energies across spin channels, we find that the coupling is not symmetric; for the spin-up orbitals, the self-energy is noticeably weaker for $|\omega|\le 120~\text{meV}$ compared to its spin-down counterpart.
Similarly, for the $a_1$ manifold, the coupling near $\omega=60$~meV is more pronounced for the spin-up orbital.
This spin-selectivity affects both the position and relative intensity of the vibronic sidebands, demonstrating that accurate spectral functions require not only mode- and momentum-resolved, but also spin-resolved $e$-ph matrix elements.

Our results show that the vibronic structure of the $\mathrm{NV}^-$ center is controlled by state-selective coupling to different parts of the full phonon spectrum. The $e_x/e_y$ and $a_1$ defect orbitals interact with qualitatively different phonon modes, showing distinct Fan-Migdal self-energies. In particular, the high-energy localized vibration near 160~meV and the lower-energy mixed localized/extended channel near 60~meV contribute differently depending on the orbital and spin channel.

We now turn to comparing our results to previous studies using Monte Carlo (MC) sampling~\cite{kunduQuantum2024}.
The overall widths of the MC and cumulant distributions for the $a_1^\uparrow$ and $a_1^\downarrow$ are comparable: 96~meV and 121~meV with MC, respectively, and 98~meV and 125~meV for the spectral function computed in this work.
This agreement can be understood in simple terms if we look at the independent boson model,
\begin{equation}\label{eq:IBM}
    H = \varepsilon_c c^\dagger c + g_0 c^\dagger c(b + b^\dagger) + \omega_0 b^\dagger b\;.
\end{equation}
An exact solution of this model is available and shows that the MC energy distribution
\begin{equation}
    P(\varepsilon) = \frac{1}{\sqrt{2\pi}\,g_0}
    \exp\!\left[-\frac{(\varepsilon - \varepsilon_c)^2}{2g_0^2}\right]
\end{equation}
and the cumulant spectral function
\begin{equation}
    A_E(\omega) = 2\pi\, e^{-g_0^2/\omega_0^2}
    \sum_{l=0}^{\infty} \frac{1}{l!}\!\left(\frac{g_0^2}{\omega_0^2}\right)^{l}
    \delta\!\left(\omega - \varepsilon_c + \frac{g_0^2}{\omega_0} - l\omega_0\right)
\end{equation}
share the same second moment for all coupling strengths (see Supplemental Material~\cite{supp_mat_prb}).
Yet the lineshapes differ qualitatively. The MC distribution is smooth and approximately Gaussian, whereas the cumulant spectral function exhibits pronounced multi-phonon satellites.

The MC sampling relies on the adiabatic approximation.
The electron instantaneously adjusts to a frozen lattice distortion, and  dynamical effects, such as virtual phonon exchange during electron propagation, are neglected.
The MC approach captures a Lamb-shift renormalization~\cite{kunduQuantum2023,kunduQuantum2024,monserratVibrational2016,kundu_nanodiamond} and the second-moment broadening, but misses spectral features included in the frequency-dependent self-energy, such as phonon satellites.

The Green's function formalism naturally captures dynamical correlations, as the spectral function describes electron addition or removal accompanied by phonon excitation.
The resulting broadening reflects a coherent redistribution of spectral weight into vibrational satellites (shake-up processes), a mechanism inaccessible to both static MC approaches and the lowest-order perturbation theory~\cite{yangCombined2021,yangComputational2022}.
The experimental validation of the dynamical satellites requires spectroscopic techniques sensitive to single-particle excitations. High-resolution photoemission or scanning tunneling spectroscopies may reveal defect states not as sharp peaks, but as complex polaron-like features with significant spectral weight redistribution, even at cryogenic temperatures.
Interestingly, through convolution of single-particle spectral functions, these satellites are expected to produce structure in the phonon sidebands of the optical absorption and emission lineshapes of NV$^-$ centers.

We note that our current Green's function approach assumes harmonic phonons and band-diagonal self-energies thus ignoring anharmonicities, Jahn-Teller distortions, Debye-Waller self-energy~\cite{allenTheory1976,lihmPhononinduced2020}, and off-diagonal self-energy terms.
These effects may shift quasiparticle energies, mix the degenerate $e_x/e_y$ manifold, and redistribute spectral weight, but are not expected to remove the broad multiphonon satellite structure.
Future work will extend the present cumulant framework to include off-diagonal and nonlinear electron--phonon couplings.

In conclusion, we presented a first-principles Green's function framework to describe the effect of $e$-ph interaction on the single particle energies of  color centers, and we discussed results for the NV$^-$ center in diamond.
Our framework is based on the retarded cumulant ansatz, and enables the study of dynamical effects not included in approaches such as the Huang-Rhys theory, Monte-Carlo sampling, and density matrix perturbation theory~\cite{yangCombined2021,yangComputational2022}.
We find that  $e$-ph interactions give rise to multi-phonon satellites that broaden the defect levels by more than 400~meV. This large broadening is temperature independent from 0~K to 300~K, presumably due to the large Debye temperature of diamond.
The satellite peaks describe shake-up effects occurring during photoemission and inverse photoemission processes, and their energies and intensities provide fingerprints of mode-, momentum-, spin-, and orbital-resolved $e$-ph coupling of defect states.
Experimental measurements of these features can be used to distinguish defect types through their characteristic spectral lineshapes.
Perturbations such as strain, electric fields, or magnetic anisotropy can further split or reshape the satellite manifolds and reveal how symmetry breaking modifies $e$-ph interactions in defects~\cite{zhangDislocationdriven2025}.
 The Green's function framework presented here naturally extends to excited-state dynamics and optical spectroscopy, offering a parameter-free framework to predict dynamical properties of color centers for quantum technologies.

\begin{acknowledgments}
\noindent
This work was supported by the Midwest Integrated Center for Computational Materials (MICCoM) as part of the Computational Materials Sciences Program funded by the U.S. Department of Energy  through the Argonne National Laboratory, under contract No. DE-AC02-06CH11357, and Basic Science Research Institute Fund of South Korea, whose NRF grant number is RS-2021-NR060139, and Glocal University 30 project (Grant No. 2.0081021.03).
J.P. acknowledges support from the Chicago Prize Fellowship in Theoretical Quantum Science.
This research used resources of the National Energy Research Scientific Computing Center (NERSC), a DOE Office of
Science User Facility supported by the Office of Science of the US
Department of Energy under Contract No. DE-AC02-05CH11231 using
NERSC award ALCC-ERCAP002595. The Flatiron Institute is a division of the Simons Foundation.
\end{acknowledgments}

\vspace{-5pt}
%

\end{document}